\documentclass[
reprint,
superscriptaddress,
frontmatterverbose, 
preprintnumbers,
amsmath,amssymb,
aps,
pra
]{revtex4-2}
\usepackage{graphicx}
\usepackage{dcolumn}
\usepackage{bm}
\usepackage{hyperref}
\usepackage{nameref}
\usepackage{braket}

\begin{document}

\title{General perturbation theory for local quantum uncertainty and its formulation in the linear-response regime}

\author{A. A. Jimenez-Romero}
\author{F. Rojas}
\affiliation{Departamento de Física, Centro de Nanociencias y Nanotecnología, Universidad Nacional Autónoma de México, Apartado Postal 14, 22800 Ensenada, B.C., México}

\date{\today}
\begin{abstract}
We develop a general perturbation theory for the local quantum uncertainty (LQU),
a discord-type quantifier of nonclassicality based on the Wigner--Yanase skew
information. Starting from a perturbed density matrix $\rho = \rho_0 + \epsilon\rho_1$,
we derive an explicit first-order expansion of $\rho^{1/2}$ using an integral
representation based on the gamma function, and reduce the LQU optimization to the
diagonalization of a $(d_1^2-1) \times (d_1^2-1)$ matrix
$w = w^0 + w^1$ defined in terms of the
generators of $\mathrm{SU}(d_1)$. The resulting framework is valid for composite
systems of arbitrary dimension $d_1 \times d_2$ and provides a direct computational
route to the LQU from the spectral decomposition of the unperturbed state $\rho_0$.
We further specialize the theory to the quantum linear response regime, where the
perturbation $\rho_1$ is generated by a time-dependent external field coupled to the
system through an operator $\hat{A}$. In this regime, $w^1$ acquires an
explicit dependence on the driving frequency $\omega$, the eigenstates and occupation
probabilities of the equilibrium Hamiltonian $H_0$, and the matrix elements of
$\hat{A}$, establishing a direct link between the LQU and the spectral and dynamical
structure of the system. The formalism is applicable to a broad class of physical
models, including quantum spin arrays with anisotropic and spin-orbit interactions,
atoms coupled to cavity fields, and optomechanical systems. To illustrate the
approach, we apply the theory to the isotropic Heisenberg model of two coupled
spins subject to a local periodic magnetic field, deriving closed-form analytical
expressions for the LQU as a function of temperature $T$ and frequency $\omega$.
Comparison with the concurrence shows that, above the entanglement critical
temperature $T_c$, the external perturbation induces a resonantly enhanced quantum
discord without generating entanglement. This demonstrates that frequency acts as
a tunable modulator of nonclassicality in the linear response regime, an effect
that is purely of quantum-discord type and a property inaccessible to entanglement-based
quantifiers.
\end{abstract}

\maketitle

\section{INTRODUCTION}
\label{sec:intro}

Quantum correlations in composite systems constitute a fundamental resource in quantum information processing. At the beginning of this century, such correlations were largely considered synonymous with entanglement --- the non-separability of quantum states --- which has been a central topic of research in quantum computing. Entanglement and its quantification have been extensively studied, both theoretically and experimentally, across a broad range of physical systems \cite{horodecki2009quantum,amico2008entanglement}. It is essential to quantum information technologies, underpinning protocols such as quantum teleportation  \cite{hu2023progress}, superdense coding  \cite{williams2017superdense,dutta2023absolutely}, and quantum computation \cite{nielsen2001quantum}.In recent years, other forms of quantum correlations distinct from entanglement have been identified, collectively referred to as discord-like quantum correlations \cite{zurek2000einselection, ollivier2001quantum}, which are rooted in local measurements and quantum information theory. The fact that discord can exist without entanglement \cite{datta2008studies, datta2008quantum}  implies that it may be nonzero for separable mixed states --- making it a property that transcends non-separability and is therefore of independent quantitative interest.

The original Quantum discord in a composite system is defined as the difference between the total mutual information of the two subsystems and the classical information accessible through a local measurement on one of them \cite{ollivier2001quantum}. Such a measurement extracts purely quantum information, sometimes referred to as "quantumness." Quantum discord, initially formulated as the minimum conditional entropy after measurement \cite{henderson2001classical}, generally involves complex optimization procedures, making its direct calculation challenging. To address this difficulty, alternative measures of discord-like correlations have been introduced that are simpler to compute \cite{LGCQ}, particularly for bipartite systems described by a density matrix  $\rho_{AB}$ in the Hilbert space $\mathcal{H}_A^{d_1} \otimes \mathcal{H}_B^{d_2}$.  These include geometric measures based on distance \cite{fanchini2017lectures} and measures derived from local measurements through the minimization of quantum information functionals. Among the latter, two prominent examples are the quantum Fisher information (QFI)  \cite{liu2020quantum}, which, when optimized over all local measurements on subsystem  $A$, gives rise to the quantum interferometric power (QIP) \cite{girolami2014quantum} or local quantum Fisher information (LQFI) \cite{kim2018characterizing}; and the Wigner-Yanase skew information \cite{wigner1963information}, which, when minimized over local measurements, yields the local quantum uncertainty (LQU) \cite{girolami2013characterizing}. Both LQFI and LQU satisfy the defining criteria for quantum-discord-type quantifiers: (a) they vanish if and only if the density matrix is classical–quantum (CQ) or  classical-classical (CC) ; (b) they are invariant under unitary transformations; and (c) they are contractive under completely positive trace-preserving (CPTP) maps on the unmeasured subsystem. The LQU \cite{girolami2013characterizing} is grounded in the Wigner-Yanase skew information, defined for the bipartite state  $\rho_{AB}$ 
 and a local observable $K_A^{\Lambda}$ acting on subsystem  $A$  (with spectrum $\Lambda$ and dimension $d_1$) as  $I_w= -\frac{1}{2}tr([\rho_{AB}^{1/2},K_A^{\Lambda}\otimes \mathbb{I}_B ]^2)$ \cite{wigner1963information}. Minimization of this quantity over all local observables  $K_A^{\Lambda}$ yields a discord-type measure of nonclassicality.

The LQU has been extensively studied in recent years, with particular attention to the role of temperature in various physical models. It was originally introduced for bipartite qubit-qudit ( $2 \otimes d$ )  systems \cite{girolami2013characterizing}, for which a closed-form procedure exists to determine the relevant matrix eigenvalues. The LQU has subsequently been extended to multipartite systems \cite{ali2020local,chavez2022environment}, where appropriately chosen subsystem partitions enable the characterization of different degrees of quantumness, demonstrating the flexibility of the measure beyond the bipartite setting. Several recent studies have derived analytical expressions for the LQU in two-spin Heisenberg models at thermal equilibrium and compared them with other nonclassicality quantifiers. Fedorova and Yurischev \cite{fedorova2022behavior}  investigated the behavior of three measures of quantum correlations --- entropic quantum discord, LQU, and LQFI --- in a fully anisotropic Heisenberg model of two coupled spins, including Dzyaloshinskii–Moriya (DM) and Kaplan–Shekhtman–Entin–Wohlman–Aharony (KSEA) interactions. Aroui \textit{et al.} \cite{aroui2022characterizing} examined thermal quantum correlations in a two-spin Heisenberg XXZ model under DM and KSEA interactions, using the trace distance discord (TDD) and LQU as quantifiers. Zidan \textit{et al.} \cite{zidan2022local}  investigated the preservation of quantum correlations in a two-spin Heisenberg system subjected to a time-dependent external magnetic field, employing the LQU and QIP as quantifiers. The LQU has also been evaluated for atoms interacting locally in a cavity \cite{mohamed2021local} and for two-level atoms coupled to a thermal reservoir \cite{dahbi2022dynamics}, among other configurations. The LQU has also been applied in quantum biology: Chávez-Huerta and Rojas \cite{chavez2022environment} employed it to quantify quantum correlations in a chromophore network within the Fenna–Matthews–Olson (FMO) pigment–protein complex. By computing the LQU across various subsystem partitions --- including one-versus-six, one-versus-two, and pairwise chromophore configurations --- they identified the dominant correlation families and revealed an environment-assisted quantum discord (ENAQD) mechanism, characterized by a maximum of accumulated discord at a specific dephasing rate. 

Although the LQU has been studied predominantly in specific spin models, there remains ample opportunity to extend this framework to a broader class of systems characterized by an unperturbed density matrix  $\rho_0$ and a perturbation $\rho_1$ obtained, for instance, from perturbation theory applied to the Lindblad master equation \cite{soto2024matrix, villegas2016application,li2014perturbative} or from linear response theory defined by an equilibrium Hamiltonian $H_0$ and an external perturbation $H_1$ \cite{des1968linear}. In this work, we develop a general perturbation theory for the LQU applicable to a broad class of quantum systems. The central object is a perturbed density matrix $\rho = \rho_0 + \epsilon \rho_1$ with   $\epsilon \ll 1 $. Since the skew information involves the square root of the density matrix,  it is necessary to expand $\rho^{1/2}$; we derive this expansion explicitly using the Stieltjes representation \cite{schilling2012bernstein}.
This approach expresses $\rho^{1/2}$ in terms of the spectral decomposition of $\rho_0$. Optimizing the skew information over the local measurement operator --- taken to be any element of the SU($d_1$) algebra, corresponding to a qudit of dimension $d_1$ --- yields a general first-order expansion of the LQU for an arbitrary perturbation $\rho_1$. This optimization reduces the computation of the LQU to the diagonalization of a $(d_1^2-1) \times (d_1^2-1)$ matrix $w$.  We then specialize this framework to the quantum linear response regime, where the total Hamiltonian is  $H=H_0 + H_1$ with $H_1$ small, and the equilibrium density matrix is $\rho = \rho_0 + \epsilon \rho_1$ . The perturbation $\rho_1$ is generated by a time-dependent external field coupled to the system via  $H_1 = -f(t)\hat{A}$. The goal is to characterize how the matrix $w$ --- evaluated at thermal equilibrium --- is modified by the external perturbation. To this end, we employ the formalism of linear response theory \cite{des1968linear, stinchcombe1978kubo} to derive an explicit expression for  $\rho_1$ in terms of the spectral structure of the equilibrium state  $\rho_0 = \sum_n \lambda_{\alpha_n} \ket{\alpha_n}\bra{\alpha_n}$, the eigenvalues and eigenstates of $H_0$, and the matrix elements $\braket{\alpha_n|\hat{A}|\alpha_m}$ of the external operator. This framework provides a systematic means of tracking the effect of the external perturbation on the nonclassicality of the system, with an external periodic force $f(t)$ introducing an explicit dependence on the driving frequency $\omega$. The resulting framework is applicable to a broad range of physical systems characterized by a Hamiltonian $H = H_0 + H_1$, including quantum spin arrays, cavity QED systems, and optomechanical platforms. Its generality further permits extensions to systems with multiple degrees of freedom, such as optoelectromechanical systems. Finally, as a concrete illustration, we apply the formalism to the isotropic Heisenberg model of two coupled spins subject to a local periodic magnetic field, deriving closed-form analytical expressions for the LQU as a function of temperature and driving frequency. These results are complemented by a calculation of the concurrence, enabling a direct comparison between discord-type correlations and entanglement. 

The remainder of this article is organized as follows. In Sec.\ref{sec:LQU-PT}, we develop the general perturbation theory for the LQU in the SU($d_1$) framework, including the expansion of $\rho^{1/2}$ and its explicit reduction to the $2\otimes d$ case. In Sec.  \ref{sec:LQU-LR}, we derive the perturbed density matrix within the linear response regime and obtain the corresponding expressions for the matrix $w$. In Sec. \ref{sec:LQU-Heisenberg-model}, we illustrate the approach by applying  the formalism to the isotropic Heisenberg model of two coupled spins, presenting closed-form expressions for the LQU as functions of temperature and frequency and comparing them with the entanglement with the concurrence.

 \section{EXPANSION FOR LOCAL QUANTUM UNCERTAINTY IN PERTURBATION THEORY}
 \label{sec:LQU-PT}

The LQU is a discord-type measure of quantum correlations that isolates the genuinely quantum contribution to measurement uncertainty. It is based on the Wigner–Yanase skew information Ref.~\cite{wigner1963information}, defined for a density operator $\rho$ and an observable $K$ as 

\begin{equation}
I_w(\rho,K):= - \frac{1}{2}\operatorname{tr}\{[\rho^{1/2},K]^2\}.
\label{eq:I_w_definition}
\end{equation}

Expanding the commutator yields

\begin{equation}
= tr(\rho K^2)-tr(\rho^{1/2}K\rho^{1/2}K).
\end{equation}

The LQU is obtained by minimizing the skew information over all local measurements performed on one of the subsystems. This minimization is meaningful only when  $K$ is a local operator, for a bipartite system composed of subsystems $A$ and $B$, such operators take the form  $\lbrace K_A^{\Lambda}:= K_A^{\Lambda} \otimes \mathbb{I}_B  \rbrace$, where $K_A^{\Lambda}$ is a set of nondegenerate Hermitian operators acting on subsystem $A$ with spectrum $\Lambda$. The LQU of subsystem $A$  is therefore defined as

\begin{equation}
 U_A^{\Lambda} (\rho_{AB}):=min_{K^{\Lambda}} I(\rho_{AB}, K_A^{\Lambda} \otimes \mathbb{I}_B ).
\end{equation}

We now introduce a perturbative expansion of the density matrix, $\rho = \rho_0 + \epsilon \rho_1$, and write the corresponding expansion of its square root as $\rho^{1/2} = \rho_0^{1/2} + \epsilon \rho_1^e$. The explicit derivation of $\rho_1^e$ is presented in Sec.~\ref{sec:rho^1/2} . For notational convenience, we henceforth denote $K_A^{\Lambda} \otimes \mathbb{I}_B$ simply as $K$. Substituting this expansion into the skew information and retaining terms to first order in $\epsilon$, we obtain

\begin{equation}
I_w = tr[(\rho_0 + \rho_1)K^2] - tr[(\rho_0^{1/2} + \epsilon\rho_1^e)K(\rho_0^{1/2} + \epsilon\rho_1^e)K],
\end{equation}

expanding the expression to first order in $\epsilon$, we have

\begin{equation}
\begin{split}
I_w = I_w^0 + I_w^1 
\end{split}
\label{eq:Iw_initial}
\end{equation}

were

\begin{equation}
I_w^0= tr[\rho_0 K^2] - tr[\rho_0^{1/2} K \rho_0^{1/2}K]
\label{eq:Iw0_initial}
\end{equation}

is the skew information associated with the equilibrium state, and

\begin{equation}
I_w^1 = \epsilon \{ tr[\rho_1 K^2]  -2 tr[ \rho_0^{1/2} K \rho_1^e K] \}
\label{eq:Iw1_initial}
\end{equation}

represents the contribution due to the perturbation.

To optimize the skew information over all local observables on subsystem 
$A$, we adopt the  SU($d_1$) framework \cite{closedformwang} and parametrize the local operator as  $K= n \cdot T \otimes \mathbb{I}_{d_2}$  where $n =(n_1,n_2,\ldots,n_{d_1^2-1})$ is a unit vector( $|n|=1$) and $T = (T_1, T_2 \ldots,T_{d_1^2-1})^T$ collects the generators of SU( SU($d_1$)). These generators satisfy the algebra $T_i T_j=\mathrm{i} \sum_k f_{i j k} T_k+\sum_k g_{i j k} T_k+\frac{2}{d} \delta_{i j} \mathbb{I}_d$, where the symmetric and antisymmetric structure constants are $f_{i j k}=\frac{1}{4 \mathrm{i}} \operatorname{Tr}\left(\left[T_i, T_j\right] T_k\right)$ and $ g_{i j k}=\frac{1}{4} \operatorname{Tr}\left(\left\{T_i, T_j\right\} T_k\right)$, respectively.
Expanding $K^2$ in this basis as  $(n \cdot T \otimes \mathbb{I}_{d_2})^2 = \sum_{ij}n_i n_j (T_i T_j \otimes \mathbb{I}_{d_2})$ and introducing the generalized Bloch vector $L$ of the subsistem $A$

\begin{equation}
\begin{aligned} & L=\left(\operatorname{Tr}\left(\rho T_1 \otimes \mathbb{I}_{d_2}\right), \ldots, \operatorname{Tr}\left(\rho T_k \otimes \mathbb{I}_{d_2}\right), \ldots, \right.  \\ &
\left. \operatorname{Tr}\left(\rho T_{d_1^2-1} \otimes \mathbb{I}_{d_2}\right)\right)^T \\& \end{aligned}
\label{eq:vector_L}
\end{equation}

whose components are the expectation values of the SU($d_1$) generators with respect to $\rho$,  we adopt the compact notation of \cite{closedformwang}

\begin{equation}
\begin{aligned} & F_{i j}=\left(f_{i j 1}, \ldots, f_{i j k}, \ldots, f_{i j d_1^2-1}\right) \\  \\ & G_{i j}=\left(g_{i j 1}, \ldots, g_{i j k}, \ldots, g_{i j f_1^2-1}\right). \\ \\ & \end{aligned}
\label{eq:vectors_F_and_G}
\end{equation}

Substituting the definitions in Eqs.~(\ref{eq:vector_L},\ref{eq:vectors_F_and_G}) into Eqs.~(\ref{eq:Iw0_initial}) and (\ref{eq:Iw1_initial}), one obtains the equilibrium contribution 

\begin{equation}
\operatorname{Tr}[\rho_0\left(n \cdot T \otimes \mathbb{I}_{d_2}\right)^2]=\sum_{i, j} n_i n_j\left[\left(\mathrm{i} F_{i j}+G_{i j}\right) L\right]+\frac{2}{d_1},
\end{equation}

and the perturbative contribution

\begin{equation}
    \begin{split}
        \operatorname{Tr}[\rho_1\left(n \cdot T \otimes \mathbb{I}_{d_2}\right)^2] & = \sum_{ijk}n_i n_j [(iF_{ij} + G_{ij}) tr(\rho_1 T_k \otimes \mathbb{I}_{d_2})] \\
         &+ \frac{2}{d_1}tr(\rho_1 \delta_{ij} \mathbb{I}_{d_1}) \\
         &= \sum_{ij} n_i n_j [iF_{ij} + G_{ij}]L_1
    \end{split}
\end{equation}  

where 

\begin{equation}
    \begin{split}
        L^1= &\left( \operatorname{Tr}\left(\rho_1 T_1 \otimes \mathbb{I}_{d_2}\right), \cdots,  \operatorname{Tr}\left(\rho_1 T_k \otimes \mathbb{I}_{d_2}\right), \right.\\
        &\left. \cdots, \operatorname{Tr}\left(\rho_1 T_{d_1^2-1} \otimes \mathbb{I}_{d_2}\right)\right)^T. 
    \end{split}
\label{eq:vector_L1}
\end{equation}

is defined analogously to $L$, but now using the perturbed density matrix $\rho_1$.

For the last two terms in Eqs.~(\ref{eq:Iw0_initial}) and (\ref{eq:Iw1_initial}) involving $\rho_0^{1/2}$ and $\rho_1^e$ we have 
\begin{equation}
 tr[\rho_0^{1/2} K \rho_0^{1/2} K] = \sum_{i, j} n_i n_j tr[\rho_0^{1/2} (T_i \otimes \mathbb{I}_{d_2}) \rho_0^{1/2} (T_j \otimes \mathbb{I}_{d_2})]
\label{eq:trace1_initial}
\end{equation}

for the equilibrium state, and

\begin{equation}
 tr[\rho_0^{1/2} K \rho_1^{e} K] = \sum_{i, j} n_i n_j tr[\rho_0^{1/2} (T_i \otimes \mathbb{I}_{d_2}) \rho_1^e (T_j \otimes \mathbb{I}_{d_2})]
 \label{eq:trace2_initial}
\end{equation}

for the perturbative element  of $\rho^{1/2}$  denoted as $\rho_1^e $ .
\\
By using the antisymmetric property  $F_{ij} = - F_{ji}$, Ref.~\cite{bossion2021general}, and incorporating the previous results into Eq.~(\ref{eq:Iw_initial}), yields the complete first-order expression for the skew information

\begin{equation}
\begin{split}
I_w = \sum_{ij} n_i n_j  \{ [G_{ij}L + \epsilon G_{ij}L^1] \\ -  tr[\rho_0^{1/2} (T_i \otimes  \mathbb{I}_{d_2}) \rho_0^{1/2} (T_j \otimes  \mathbb{I}_{d_2})]\\
 - 2\epsilon tr[\rho_0^{1/2} (T_i \otimes  \mathbb{I}_{d_2}) \rho_1^e (T_j \otimes  \mathbb{I}_{d_2})]\} + \frac{2}{d_1}.
\end{split}
\label{eq:Iw_expression_initial}
\end{equation}

Minimizing over $K= n \cdot T\otimes \mathbb{I}_{d_2}$ then reduces the LQU to the problem of finding the maximum eigenvalue of the $(d_1^2-1)\times(d_1^2-1)$  matrix $w = w^0 + w^1$

\begin{equation}
u_a =  \frac{2}{d_1} - max(w)
\end{equation}

where $max(w)$ denotes the maximum eigenvalue of the $(d_1^2 -1) \times (d_1^2 -1)$  matrix 

\begin{equation}
w = w_{ij}^0 + w_{ij}^1
\label{eq:w0_+_w1_initial}
\end{equation}

with elements corresponding to the equilibrium state $\rho_0$

\begin{equation}
w_{ij}^0 =   tr[\rho_0^{1/2} ( T_i \otimes \mathbb{I}_{d_2}) \rho_0^{1/2} ( T_j \otimes  \mathbb{I}_{d_2})] - G_{ij}L(\rho_0),
\label{eq:wij0_general}
\end{equation}

and to the perturbed state $\rho_1$

\begin{equation}
 w_{ij}^1 = 2\epsilon tr[\rho_0^{1/2} (T_i \otimes  \mathbb{I}_{d_2}) \rho_1^e (T_j \otimes  \mathbb{I}_{d_2})] - \epsilon G_{ij}L^1(\rho_1).
 \label{eq:wij1_general}
\end{equation}

In the limit $\epsilon \to 0$, the perturbative correction vanishes $w \to w^0$, recovering the known result of Ref.~\cite{closedformwang}. Eqs.~(\ref{eq:wij0_general}) and (\ref{eq:wij1_general}) constitute the central result of this section, providing a general and computationally direct procedure for evaluating the LQU in composite systems subject to an arbitrary perturbation $\rho_1$.

\subsection{First-order expansion of $\rho^{1/2}$ via the gamma function representation}
\label{sec:rho^1/2}

To obtain explicit expressions for the matrices  $w_{ij}^0$ and $w_{ij}^1$,  it is necessary to first derive a perturbative expansion of $\rho^{1/2}$. To this end, we derive a systematic expansion of a perturbed density matrix of the form $\rho = \rho_0 + \epsilon \rho_1$ based on the integral representation of the square root via the Gamma and Beta functions \cite{choi2009integral}, which leads naturally to the Stieltjes representation. For clarity of notation, we temporarily denote $\rho \rightarrow \gamma$ , and write the Stieltjes integral representation of $\gamma^{1/2}$ \cite{schilling2012bernstein} as

\begin{equation}
\gamma^{1/2} = \frac{sen(\pi/2)}{\pi} \int_0^{\infty} \frac{\gamma}{\gamma + t}t^{-1/2}dt .
\label{eq:gamma_1/2_integral}
\end{equation}

A Taylor-like expansion  \cite{daletskii1965integration}  of  $\gamma^{1/2}$  is then constructed as 

\begin{equation}
\gamma^{1/2}= \gamma_0^{1/2} + \epsilon \left. \frac{d}{d\epsilon}\right|_{\epsilon=0} \gamma^{1/2}  + O(\epsilon^2),
\label{eq:gamma_expansion}
\end{equation}

where the first-order term is determined via the operator derivative formalism \cite{higham2008functions, higham2014higher}. Denoting the perturbation as

\begin{equation}
\epsilon \rho_1 \equiv \nu
\end{equation}

and introducing the operators

\begin{equation}
\gamma_+ = (\rho_0 + \epsilon \nu) ;\hspace{.5cm} \gamma_0 = \rho_0,
\label{eq:gamma0_gamma+}
\end{equation}

the first-order derivative takes the form

\begin{equation}
\left. \frac{d}{d\epsilon}\right|_{\epsilon=0} \gamma^{1/2}  = \lim_{\epsilon \rightarrow 0} \frac{\gamma_+^{1/2} - \gamma_0^{1/2}}{\epsilon}.
\label{eq:derivative_first_order}
\end{equation}

Substituting  $\gamma_0^{1/2}$ from Eq.~(\ref{eq:gamma0_gamma+})  and the integral representation of  $\gamma_+^{1/2}$ (\ref{eq:gamma_1/2_integral})  on the right-hand side of  from Eq.~ (\ref{eq:gamma_1/2_integral}) into the right-hand side of Eq.~(\ref{eq:derivative_first_order}), and inserting the result into Eq.~(\ref{eq:gamma_expansion}), one arrives at the following integral expression for $\rho^{1/2}$

\begin{equation}
\begin{split}
\rho^{1/2}= \rho_0^{1/2} + \lim_{\epsilon \rightarrow 0} \frac{1}{\epsilon \pi} \int\frac{\epsilon   [\nu, \rho_0] + \epsilon \nu t}{(\rho_0 + \epsilon \nu + t)(\rho_0 + t)}t^{-1/2}dt \\
+ O(\epsilon^2).      
\end{split}
\label{eq:rho_1/2_integral}
\end{equation}

The detailed  evaluation of Eq.~(\ref{eq:rho_1/2_integral}) is provided in Appendix~\ref{apen:rho^1/2}. Expressing the unperturbed density operator in its spectral decomposition, $\rho_0 = \sum_i \lambda_i \ket{\psi_i}\bra{\psi_i}$, and projecting onto the eigenstates$\ket{\psi_i}$ of $\rho_0$, the integrals in Eq.~(\ref{eq:rho_1/2_integral}) can be evaluated in closed form, yielding 

\begin{equation}
\rho^{1/2} = \sum_{i}\lambda_{i}^{1/2}\ket{\psi_i}\bra{\psi_i} + \epsilon\sum_{i j}\rho_{1_{ij}}^e \ket{\psi_i}\bra{\psi_j} + \epsilon^2(O^2)
\label{eq:rho1/2_general}
\end{equation}

 where the matrix elements of $\rho_{1_{ij}}^e$ are given by 

\begin{equation}
\rho_{1_{ij}}^e = \rho_{1_{ij}}
 \left( \frac{\lambda^{1/2}_{i} -\lambda^{1/2}_{j}}{\lambda_{i} -\lambda_{j}} \right).
 \label{eq:rho1_e}
\end{equation}

Equation~(\ref{eq:rho1/2_general}), of the form $\rho^{1/2} = \rho_0^{1/2} + \epsilon\rho^{e}_1$, constitutes the key ingredient for evaluating the skew information and the LQU within the perturbative framework developed in the following subsections.

\subsection{General expressions for $w_{ij}^0$ and $w_{ij}^1$ in the spectral decomposition of $\rho_0$}

The spectral decomposition of  $\rho_0$ provides a natural and computationally convenient basis in which to evaluate the matrices $w_{ij}^0$  and $w_{ij}^1$, since all quantities of interest can be expressed directly in terms of the eigenvalues and eigenstates of the unperturbed state. Substituting the explicit form of $\rho_1^e$  from Eq. (\ref{eq:rho1_e}), together with the spectral decomposition of  $\rho_0$, into Eqs.~(\ref{eq:wij0_general}) and (\ref{eq:wij1_general}), we obtain the components of the generalized Bloch vectors $L$ and $L^1$. The vector $L$ , which characterizes the unperturbed state  $\rho_0$ , is given by

\begin{equation}
    \begin{split}
        L = & \left(\sum_{n} \lambda_{n} \braket{\psi_n| T_1 \otimes \mathbb{I}_{d_2}|\psi_n}, \ldots,\right. \\
&\sum_{n} \lambda_{n} \braket{\psi_n| T_k \otimes  \mathbb{I}_{d_2}|\psi_n}, \ldots \\
&\left.\sum_{n} \lambda_{n} \braket{\psi_n| T_{d^2-1} \otimes  \mathbb{I}_{d_2}|\psi_n} \right)^T
    \end{split}
\end{equation}

 while $L^1$, which encodes the first-order correction due to $\rho_1$, takes the form

\begin{equation}
     \begin{split}
         L^1 = &\left( \sum_{nm} \rho_{1_{nm}} \braket{\psi_m|T_1\otimes \mathbb{I}_{d_2}|\psi_n},\ldots, \right. \\
        &\sum_{nm} \rho_{1_{nm}} \braket{\psi_m|T_k\otimes \mathbb{I}_{d_2}|\psi_n},\ldots,\\
        &\left. \sum_{nm} \rho_{1_{nm}} \braket{\psi_m|T_{d^2-1}\otimes \mathbb{I}_{d_2}|\psi_n} \right)^T.
     \end{split}
\end{equation}

Substituting these expressions into the trace terms of.~(\ref{eq:trace1_initial}) and (\ref{eq:trace2_initial}), evaluated in the eigenbasis of $\rho_0$, yields the explicit forms of the matrices $w_{ij}^0$ and $w_{ij}^1$  given by

\begin{equation}
    \begin{split}
        w_{ij}^0 = &\sum_{nm}\lambda_n^{1/2}\lambda_m^{1/2} \braket{\psi_n|T_i \otimes \mathbb{I}_{d_2}|\psi_m}\braket{\psi_m|T_j \otimes \mathbb{I}_{d_2}|\psi_n}\\
        &- G_{nm}L
    \end{split}
\label{eq:w0_basis}
\end{equation}

and

\begin{equation}
    \begin{split}
        w_{ij}^1 = & \sum_{nml}\lambda_n \frac{(\lambda^{1/2}_m-\lambda^{1/2}_l)}{\lambda_m-\lambda_l} \rho_{1_{ml}} \\
        &\times \braket{\psi_n|T_i \otimes \mathbb{I}_{d_2}|\psi_m}\braket{\psi_l|T_j \otimes \mathbb{I}_{d_2}|\psi_n}- \epsilon G_{nm}L^1.
    \end{split}
\label{eq:w1_basis}
\end{equation}

Together, these expressions  constitute a general and direct method for evaluating the LQU from the spectral data of the unperturbed state $\rho_0$ for an arbitrary perturbation $\rho_1$ expanded in the eigenbasis of $H_0$.

\subsection{Reduction to qubit–qudit ( $2\otimes d$) systems}

The most widely studied setting for evaluating the LQU is one in which one subsystem is a qubit, i.e., a two-level system. This framework encompasses a broad class of physical models, including spin-spin systems, two-level atoms interacting with light fields (e.g., Rabi models), and systems involving mechanical degrees of freedom.

Setting $d_1=2$ in Eq.~(\ref{eq:w0_+_w1_initial}), the SU(2) structure constants satisfy $g_{ijk}=0$, and consequently $G_{ij}=0$ \cite{bossion2021general}. The closed-form expression for the LQU in $2 \otimes d$ quantum systems  \cite{girolami2013characterizing} then reduces to

\begin{equation}
 u_A^{\Lambda} (\rho_{AB}) = 1 - max\{w_{AB}\}
\label{eq:LQU_general_2xd}
\end{equation}

where $w_{AB} = w^0_{ij} + w^1_{ij}$  is a  $3\times 3$ matrix  whose elements are given by the unperturbed contribution

\begin{equation}
w^0_{ij} =\operatorname{tr}\left\{\rho_{0}^{1 / 2}\left(\sigma_{i A} \otimes \mathbb{I}_B\right) \rho_{0}^{1 / 2}\left(\sigma_{j A} \otimes  \mathbb{I}_B\right)\right\}
\label{eq:wij0_general2xd}
\end{equation}

and the first-order perturbative correction

\begin{equation}
w^1_{ij}=2\epsilon \operatorname{tr}\left\{\rho_0^{1 / 2}\left(\sigma_{i A} \otimes \mathbb{I}_B\right) \rho_1^{e}\left(\sigma_{j A} \otimes  \mathbb{I}_B\right)\right\}.
\label{eq:wij1_2xd_general}
\end{equation}

Substituting the spectral decomposition of $\rho_0$  into Eqs.~(\ref{eq:wij0_general2xd}) and (\ref{eq:wij1_2xd_general}), the explicit expressions in the eigenbasis of  $\rho_0$ read

\begin{equation*}
w_{ij}^0 = \sum_{nm}\lambda_n^{1/2}\lambda_m^{1/2} \braket{\psi_n|\sigma_{i A} \otimes  \mathbb{I}_B|\psi_m}\braket{\psi_m|\sigma_{j A} \otimes  \mathbb{I}_B|\psi_n}
\label{eq:wij0_general_2xd_basis}
\end{equation*}

and

\begin{equation}
\begin{split}
w_{ij}^1 = & \sum_{nml}\lambda_n \frac{(\lambda^{1/2}_m-\lambda^{1/2}_l)}{\lambda_m-\lambda_l} \rho_{1_{ml}} \\ 
&\times \braket{\psi_n|\sigma_{i A} \otimes  \mathbb{I}_B|\psi_m}\braket{\psi_l|\sigma_{j A} \otimes  \mathbb{I}_B|\psi_n},
\end{split}
\label{eq:wij1_2xd_general_basis}
\end{equation}

where $\sigma_i$ denote the Pauli matrices, with indices $i, j \in \{x, y, z\}$. In the limit $\varepsilon \to 0$, the perturbative correction vanishes and one recovers the unperturbed result $\left(w_{A B}\right)_{i j}=\operatorname{tr}\left\{\rho_{A B}^{1 / 2}\left(\sigma_{i A} \otimes \mathbb{I}_B\right) \rho_{A B}^{1 / 2}\left(\sigma_{j A} \otimes  \mathbb{I}_B\right)\right\}$  of Ref.~\cite{girolami2013characterizing}. Furthermore, it can be shown that for pure states, Eq.~(\ref{eq:LQU_general_2xd}) reduces to the linear entropy of entanglement, which is directly related to the concurrence \cite{girolami2013characterizing}. The formulation presented here thus provides a systematic and computationally convenient framework for evaluating the LQU in bipartite systems where subsystem $A$ carries an SU(2) structure.

\section{LOCAL QUANTUM UNCERTAINTY IN THE  QUANTUM LINEAR RESPONSE REGIME}
\label{sec:LQU-LR}

In this section, we apply the general perturbation framework developed in Sec.~\ref{sec:LQU-PT} to the physically relevant case in which the perturbed density matrix $\rho = \rho_0 + \epsilon \rho_1$ is generated by an external time-dependent field. Specifically, $\rho_1$ is derived within the quantum linear response regime, which provides an explicit expression for the first-order correction to the equilibrium state in terms of the spectral properties of $H_0$ and the coupling operator $\hat{A}$.

\subsection{Linear response formalism}

To study how the expectation value of a physical observable responds to an external force $f(t)$, it is necessary to introduce a function relating these two quantities; this object is known as the response function. Linear response theory provides a systematic framework valid in the regime where the change in any observable is linear in the applied force $f(t)$ \cite{des1968linear, stinchcombe1978kubo}.

We consider a system governed by an unperturbed Hamiltonian $H_0$  and initially prepared in thermal equilibrium, so that its state is described by the canonical density matrix  

\begin{equation}
\rho_0 = \frac{e^{-\beta H_0}}{Z}
\label{eq:rho_canonical_LR}
\end{equation}

where $Z= \operatorname{tr} \{e^{-\beta H_0}\}$  is the partition function and $\beta= \frac{1}{kT}$ is the inverse temperature. When an external force $f(t)$ is applied, the total Hamiltonian acquires a time-dependent perturbation

\begin{equation}
H= H_0 + H_1
\label{eq:total_H_LR}
\end{equation}

where

\begin{equation}
H_1= -\hat{A}f(t)
\label{eq:H1_LR}
\end{equation}

and is the operator through which the system couples to the external field.

Since $\rho_0$ commutes with $H_0$ by construction, i.e., $\quad\left[H_0, \rho_0\right]=0$, the full time-dependent density matrix can be decomposed as  

\begin{equation}
\rho(t)=\rho_0+\rho_1(t) 
\end{equation}

where $\rho_1$  denotes the first-order correction induced by the perturbation $H_1$.
Solving the von Neumann equation to first order in $H_1$ yields \cite{des1968linear}

\begin{equation}
\rho(t)=\rho_0+i \int_{-\infty}^t\left[\hat{A}\left(t^{\prime}-t\right), \rho_0\right] f\left(t^{\prime}\right) d t^{\prime} .
\label{eq:integral_density_matriz_LR}
\end{equation}

This integral expression, which encodes the generalized susceptibility of the system, serves as the starting point for deriving the explicit matrix form of $\rho_1$ used in the subsequent evaluation of the LQU.

\subsection{First-order correction to the density matrix}
\label{subsec:density_matrix_LR}

Our goal is to obtain an explicit matrix representation of $\rho(t)$ in terms of the eigenstates of $H_0$  and the matrix elements of the coupling operator $\hat{A}$. We start from the integral expression in  Eq.~(\ref{eq:integral_density_matriz_LR}), where  is written in the interaction picture as

\begin{equation}
\hat{A}_0(t)=e^{i H_0 t} \hat{A} e^{-i H_0 t}.
\label{eq:A0_LR}
\end{equation}

The eigenstates and eigenvalues of  $H_0$ satisfy

\begin{equation}
 \quad H_0\left|\alpha_n\right\rangle=E_{\alpha_n}\left|\alpha_n\right\rangle
 \label{eq:H0_basis_LR}
\end{equation}

 and the equilibrium density matrix takes the spectral form  $\rho_0 = \lambda_{\alpha_n} \ket{\alpha_n}\bra{\alpha_n} $, with thermal weights  $\lambda_{\alpha_n} = \frac{e^{-\beta E_{\alpha_n}}}{Z}$ as defined in Eq. (\ref{eq:rho_canonical_LR}).  The external force is taken to be real and harmonic, 

\begin{equation}
f\left(t^{\prime}\right)=f_0 \cos \left(\omega^{\prime} t^{\prime}\right)=\frac{1}{2} f_0\left(e^{i \omega^{\prime} t^{\prime}}+e^{-i \omega^{\prime} t^{\prime}}\right) 
\end{equation}

with $f_0 \in  \mathbb{R}$.

Projecting Eq. (\ref{eq:integral_density_matriz_LR})  onto the eigenbasis  $\ket{\alpha_n}$

\begin{equation}
   \begin{split}
       \sum_{\alpha_m \alpha_{n}} \rho_{\alpha_m \alpha_{n}}(t) \ket{\alpha_{n}}\bra{\alpha_m}  =  \sum_{\alpha_m \alpha_{n}} \ket{\alpha_{n}}\bra{\alpha_{n}}\rho_0 \ket{\alpha_m}\bra{\alpha_m} \\
       + \frac{if_0}{2}\sum_{\alpha_m \alpha_{n}}\int_{- \infty}^{t} \ket{\alpha_{n}}\bra{\alpha_{n}} \left[ \hat{A}_0(t^{\prime}-t),\rho_0 \right]\\
\times (e^{i\omega't^{\prime}}+ e^{-i\omega't^{\prime}}) \ket{\alpha_m}\bra{\alpha_m}dt^{\prime},
   \end{split} 
\end{equation}

expanding the commutator, and evaluating the resulting integrals yields the matrix representation of the density operator

\begin{equation}
\rho(\omega) = \sum_{\alpha_m}\lambda_{\alpha_m}\ket{\alpha_m}\bra{\alpha_m} + \epsilon \sum_{\alpha_n \alpha_m} \rho_{1\alpha_n \alpha_m}\ket{\alpha_{n}}\bra{\alpha_m} 
\label{eq:rho_expansion_basis_LR}
\end{equation}

where

\begin{equation}
\begin{split}
 \rho_{1\alpha_n \alpha_m}= 2\pi f_0 \bra{\alpha_{n}}\hat{A}\ket{\alpha_m} (\lambda_{\alpha_m}-\lambda_{\alpha_{n}})\\
 \times F_{\alpha_n \alpha_m}(\omega)
 \end{split}
 \label{eq:rho1_expansion_basis_LR}
\end{equation}

and 

\begin{equation}
\begin{split}
F_{\alpha_n \alpha_m}(\omega) = & \left(\frac{1}{E_{\alpha_{n}} -E_{\alpha_m}-\omega -i\delta} \right. \\
& \left. + \frac{1}{E_{\alpha_{n}} -E_{\alpha_m}+ \omega -i\delta}\right).
\end{split}
\label{eq:Fnm_LR}
\end{equation}

 The resulting expression,  $\rho(\omega) = \rho_0 +  \epsilon \rho_1(\hat{A})$, separates the density matrix into its equilibrium part and a first-order correction that depends explicitly on the driving frequency  $\omega$ and the matrix elements of $\hat{A}$ in the $H_0$ eigenbasis. This structure provides the explicit form of $\rho_1$ required to evaluate the matrices $w_{ij}^0$ and $w_{ij}^1$ within the linear response regime, as carried out in the following subsection.

\subsection{LQU matrices in the driven system}

Building on the results of Secs. \ref{sec:LQU-PT} and  \ref{subsec:density_matrix_LR}, we now evaluate the explicit forms of $w_{ij}^0$ and $w_{ij}^1$ within the linear response regime. To this end, we substitute the expression for $\rho_1$ given in Eq. (\ref{eq:rho1_expansion_basis_LR}) together with the expansion of $\rho^{1/2}$ from Eq. (\ref{eq:rho1_e}) into Eq. (\ref{eq:rho_1^e_LR}), obtaining

\begin{equation}
\begin{split}
\rho_1^e =  2\pi f_0\sum_{\alpha_m \alpha_{n}} \bra{\alpha_{n}}\hat{A}\ket{\alpha_m} (\lambda_{\alpha_m}-\lambda_{\alpha_{n}})
\\
\times F_{\alpha_n \alpha_m}(\omega)  \left( \frac{\lambda^{1/2}_{\alpha_{n}} -\lambda^{1/2}_{\alpha_{m}}}{\lambda_{\alpha_{n}} -\lambda_{\alpha_{m}}} \right).
\end{split}
\label{eq:rho_1^e_LR}
\end{equation}

Computing the traces appearing in the generalized Bloch vectors $L$ and $L_1$, (\ref{eq:vector_L}) (\ref{eq:vector_L1}), as well as those  entering Eqs.~(\ref{eq:trace1_initial}) and (\ref{eq:trace2_initial}) yields the equilibrium contribution

\begin{equation}
    \begin{split}
        & tr[\rho_0^{1/2} (T_i \otimes  \mathbb{I}_{d_2}) \rho_0^{1/2} (T_j \otimes  \mathbb{I}_{d_2})] \\
        & = \sum_{\alpha_n  \alpha_m} \lambda^{1/2}_{\alpha_n} \lambda^{1/2}_{\alpha_m} \braket{\alpha_n|T_i \otimes  \mathbb{I}_{d_2} |\alpha_m} \braket{\alpha_m|T_j \otimes  \mathbb{I}_{d_2} |\alpha_n},
    \end{split}
\end{equation}

and the perturbed contribution 

\begin{equation}
    \begin{split}
        & tr[\rho_0^{1/2} (T_i \otimes \mathbb{I}_{d_2}) \rho_1^e (T_j \otimes  \mathbb{I}_{d_2})] \\
        & = - 2\pi f_0 \sum_{ \alpha_n \alpha_m \alpha_l} \lambda^{1/2}_{\alpha_n}(\lambda^{1/2}_{\alpha_m}- \lambda^{1/2}_{\alpha_l})F_{\alpha_n \alpha_m}(\omega)  \\
        & \times \braket{\alpha_m|\hat{A}|\alpha_l} \braket{\alpha_n|T_i \otimes  \mathbb{I}_{d_2} |\alpha_m} \braket{\alpha_l|T_j \otimes  \mathbb{I}_{d_2} |\alpha_n}
    \end{split}
\end{equation}

together with the explicit components  of  

\begin{equation}
    \begin{split}
       L = & \left(\sum_{\alpha_n} \lambda_{\alpha_n} \braket{\alpha_n| T_1 \otimes \mathbb{I}_{d_2}|\alpha_n}, ...,\right. \\
       & \sum_{\alpha_n} \lambda_{\alpha_n} \braket{\alpha_n| T_k \otimes  \mathbb{I}_{d_2}|\alpha_n},... \\
       &\left.\sum_{\alpha_n} \lambda_{\alpha_n} \braket{\alpha_n| T_{d^2-1} \otimes  \mathbb{I}_{d_2}|\alpha_n} \right)^T
    \end{split}
\end{equation}

and 

\begin{equation}
\begin{split}
    L^1 = & \left( \sum_{\alpha_n \alpha_m} (\lambda_{\alpha_m}- \lambda_{\alpha_n})F_{\alpha_n \alpha_m}(\omega) \right. 
\braket{\alpha_m|T_1 \otimes \mathbb{I}_{d_2} |\alpha_n},\\ 
&... \sum_{\alpha_n \alpha_m} (\lambda_{\alpha_m}- \lambda_{\alpha_n})F_{\alpha_n \alpha_m}(\omega) \braket{\alpha_m|T_k \otimes  \mathbb{I}_{d_2} |\alpha_n}, \\ 
&...\sum_{\alpha_n \alpha_m} (\lambda_{\alpha_m}- \lambda_{\alpha_n})F_{\alpha_n \alpha_m}(\omega)  \left. \braket{\alpha_m|T_{d^2-1} \otimes  \mathbb{I}_{d_2} |\alpha_n} \right)^T.
\end{split}
\end{equation}

Substituting these results into Eqs. (\ref{eq:wij0_general}) and (\ref{eq:wij1_general}), we obtain the matrices $w_{ij}^0$ and $w_{ij}^1$  expressed in terms of the eigenenergies $E_{\alpha_n}$ and eigenstates of $H_0$, the thermal occupation probabilities  $\lambda_{\alpha_n}$ of $\rho_0$, the driving frequency $\omega$, and the matrix elements of the coupling operator $\hat{A}$  

\begin{equation}
    \begin{split}
        w_{ij}^0 = & \sum_{\alpha_n  \alpha_m} \lambda^{1/2}_{\alpha_n} \lambda^{1/2}_{\alpha_m} \braket{\alpha_n|T_i \otimes  \mathbb{I}_{d_2} |\alpha_m} \braket{\alpha_m|T_j \otimes \mathbb{I}_{d_2} |\alpha_n} \\
        &- G_{ij} L
    \end{split}
\label{eq:wij0_generalRL}
\end{equation}

and

\begin{equation}
    \begin{split}
     w_{ij}^1 = & - 2 \epsilon (2\pi f_0) \sum_{ \alpha_n \alpha_m \alpha_l} \lambda^{1/2}_{\alpha_n}(\lambda^{1/2}_{\alpha_m}- \lambda^{1/2}_{\alpha_l})F_{\alpha_n \alpha_m}(\omega) \\ 
     &\times \braket{\alpha_m|\hat{A}|\alpha_l}\braket{\alpha_n|T_i \otimes \mathbb{I}_{d_2} |\alpha_m} \braket{\alpha_l|T_j \otimes \mathbb{I}_{d_2} |\alpha_n} \\
     & -\epsilon G_{ij} L^1. 
    \end{split}
\label{eq:wij1_LR}
\end{equation}

These expressions constitute the central result of this section and provide a complete computational route for evaluating the LQU in systems subject to a periodic external perturbation within the linear response regime. Their application to a specific physical model is carried out in the following section.

\section{LOCAL QUANTUM UNCERTAINTY FOR THE HEISENBERG MODEL OF TWO COUPLED SPINS}
\label{sec:LQU-Heisenberg-model}

To illustrate the approach, we apply the theoretical framework developed in Secs. \ref{sec:LQU-PT} and \ref{sec:LQU-LR} to a solvable analytical model in order to show how nonclassical properties can be characterized as functions of temperature, frequency, and coupling strength. Specifically, we consider the Heisenberg model of two coupled spins subjected to an external perturbation $\hat{A}$, for which both the LQU and the concurrence \cite{hill1997entanglement} are computed to provide a comprehensive characterization of nonclassicality. We consider the isotropic Heisenberg Hamiltonian for two coupled spins, $H_0 = J S_1\cdot S_2$ \cite{auerbach2012interacting}, where $S_1 = (\sigma_x^i,\sigma_y^i,\sigma_z^i)$, are the $\frac{1}{2}$ spin operators and $J$ is the exchange coupling constant. Working in the two-qubit computational basis $\ket{00}$, $\ket{01}$, $\ket{10}$ and $\ket{11}$,  diagonalization of $H_0$ yields four eigenstates: the singlet  $\ket{\alpha_{n=1}} = \frac{1}{\sqrt{2}}(\ket{01}-\ket{10}), $ with eigenenergy $E_{\alpha_{n=1}}= -\frac{3}{4}J$  (ground state), and the threefold-degenerate triplet states, $\ket{\alpha_{n=2}} =\frac{1}{\sqrt{2}}(\ket{01}+\ket{10})$, $\ket{\alpha_{n=3}} = \ket{11}$ and $\ket{\alpha_{n=4}} = \ket{00}$, each with eigenenergy $E_{\alpha_{n=2,3,4}} = \tfrac{1}{4}J$ separated from the singlet by an energy gap $J$.

The equilibrium density matrix is

\begin{equation}
\rho_0 = \frac{e^{-\beta J S_1 \cdot S_2}}{Z} = \sum_n \frac{\lambda_{\alpha_n}}{Z}\ket{\alpha_n}\bra{\alpha_n},
\label{eq:rho0_Heisenberg_model}
\end{equation}

where $Z= e^{3/4 \beta J} + 3e^{-1/4 \beta J}$ is the partition function and $\lambda_{\alpha_n} = \frac{e^{-\beta E_{\alpha_n}}}{Z}$ are the thermal occupation probabilities (eigenvalues of $\rho_0$). To compute the LQU in the linear response regime, we introduce an external perturbation acting locally on one subsystem via the Hamiltonian $H_1=\hat{A} f(t) = \hat{A} f_0 cos(\omega t)$, where  $\hat{A}$ is a local spin operator; specifically,
$\hat{A} \in \sigma^i_x, \sigma^i_y, \sigma^i_z$.

\subsection{ Unperturbed case ($f_0 = 0$)}

We first evaluate the LQU for the unperturbed system by specializing the  $2\otimes d$ formalism of Sec. \ref{eq:wij0_general_2xd_basis} to the present model. Expressing the matrix $w_{ij}^0$ in the eigenbasis of  $H_0$ as defined in Eq.~(\ref{eq:H0_basis_LR}), we obtain 

\begin{equation}
    w^0_{ij}=  \sum_{\alpha_n \alpha_m} \lambda_{\alpha_n}^{1/2}\lambda_{\alpha_m}^{1/2}\braket{ \alpha_n|\sigma_i \otimes \mathbb{I}_{B}|\alpha_m}\braket{ \alpha_m| \sigma_j \otimes \mathbb{I}_{B}|\alpha_n} .
\end{equation}

Evaluating the matrix elements $\braket{ \alpha_n|\sigma_i \otimes \mathbb{I}_{B}|\alpha_m}$as detailed in Appendix~\ref{apen:wij}, one finds due to the spin symmetry of the model that all off-diagonal elements of $w_{ij}^0$	vanish and that the diagonal elements are equal

\begin{equation*}
   w_{xx}^0 = w_{yy}^0 = w_{zz}^0 = 4\frac{cosh(\beta J/4)}{Z} .
\end{equation*}

Hence $w_{ij}^0$ is proportional to the identity

\begin{equation}
    w_{ij}^0 = 4\frac{cosh(\beta J/4)}{Z} \left(\begin{array}{ccc} 1 & 0 & 0 \\ 0 & 1 & 0\\ 0 & 0 & 1\end{array}\right).
\label{eq:wij0_Heisenberg_model}
\end{equation}

Substituting Eq. (\ref{eq:wij0_Heisenberg_model}) into Eq. (\ref{eq:LQU_general_2xd}) and using the explicit form of  $Z$, the LQU for the unperturbed state is

\begin{equation}
u_A = 1 - 4\frac{cosh(\beta J/4)}{Z}.
\label{eq:LQU_w0_Heisenberg_model}
\end{equation}

\subsection{Perturbed case: LQU in the presence of $H_1$}
\label{subsec:LQU_perturbed_case}

We now consider the effect of the external perturbation $H_1$.  The density matrices $\rho$, $\rho^{1/2}$ and $\rho_1$ are given by Eqs. ~(\ref{eq:rho_expansion_basis_LR}), (\ref{eq:rho1/2_general}), and (\ref{eq:rho1_expansion_basis_LR})  respectively. Specializing Eq.~(\ref{eq:wij1_LR}) to the $2 \otimes d$ case, the first-order correction to the matrix $w$ is

\begin{equation}
    \begin{split}
        w^1_{ij} =  - 2 \xi \sum_{\alpha_n \alpha_m \alpha_l} \lambda_{\alpha_n}^{1/2}(\lambda_{\alpha_m}^{1/2} - \lambda_{\alpha_l}^{1/2}) F_{\alpha_m \alpha_l}(\omega)\\
        \times \braket{\alpha_m|A|\alpha_l}\braket{ \alpha_n|\sigma_i \otimes \mathbb{I}_{B}|\alpha_m}\braket{ \alpha_l| \sigma_j \otimes \mathbb{I}_{B}|\alpha_n},
    \end{split}
    \label{eq:wij_Heisenberg_model}
\end{equation}

where $\xi = \epsilon(2\pi f_0)$ and $F_{\alpha_m \alpha_l}(\omega)$ are defined in Eq.~(\ref{eq:Fnm_LR}).

Taking  a pertubation of the type $\hat{A}= \sigma_z \otimes \mathbb{I}_{B}$, the structure of the elements $\braket{ \alpha_n|\sigma_i \otimes \mathbb{I}_{B}|\alpha_m}$ combined with the degeneracy of the triplet eigenenergies implies that all diagonal elements  $w_{ii}^1$ vanish. The only nonzero contributions arise from the off-diagonal elements with $(i,j) = (y,x)$ and $(x,y)$

\begin{equation}
    \begin{split}
        w^1_{yx} & =   -2 \xi  \sum_{\alpha_n \alpha_m \alpha_l} \lambda_{\alpha_n}^{1/2}(\lambda_{\alpha_m}^{1/2} - \lambda_{\alpha_l}^{1/2}) \braket{\alpha_m|\sigma_z \otimes \mathbb{I}_{B}|\alpha_l}\\
        &\times F_{\alpha_m \alpha_l}(\omega) \braket{ \alpha_n|\sigma_x \otimes \mathbb{I}_{B}|\alpha_m} \braket{ \alpha_l| \sigma_y \otimes \mathbb{I}_{B}|\alpha_n}\\
         & = (w^{1}_{xy})^*
    \end{split}
\end{equation}

For convenience, we introduce the shorthand

\begin{equation}
e^+ = \lambda^{1/2}_{\alpha_n}(\lambda^{1/2}_{\alpha_m}-\lambda^{1/2}_{\alpha_l}) = 2\frac{sinh(\beta J/4)}{Z},
\end{equation}
corresponding to the index combination $m=1$, $n,l \in  \{2,3,4\} $ the analogous combination with  $n,m \in  \{2,3,4\} $, $l=1$ yields $ e^-$.

Evaluating $F_{\alpha_m \alpha_l}(\omega)$  the eigenbasis of $H_0$ yields combinations of the form 
$F_{\alpha_1 \alpha_m}$ and $F_{\alpha_l \alpha_1}$. Using the antisymmetry relation of $F$

\begin{equation}
 F_{\alpha_m \alpha_l}(\omega) = - F^*_{\alpha_l \alpha_m}(\omega),
\end{equation}

these can be grouped in terms of real parts $ \operatorname{Re}[F_{\alpha_m \alpha_l}(\omega)] $, given by the spectral function

\begin{equation}
    \begin{split}
        &\operatorname{Re}[F_{\alpha_m \alpha_l}(\omega)]  \\
        & = \frac{2(E_{\alpha_m} - E_{\alpha_l})[(E_{\alpha_m} - E_{\alpha_l})^2 - \omega^2 + \delta^2]}{[(E_{\alpha_m}-E_{\alpha_l}+\omega)^2 + \delta^2][(E_{\alpha_m}-E_{\alpha_l}-\omega)^2 + \delta^2] }.
    \end{split}
    \label{eq:ReF}
\end{equation}

With the index ordering $l = 1$ and $m \in  \{2,3,4\}$ which ensures $\operatorname{Re}[F_{\alpha_m \alpha_1}(\omega)] > 0$, and noting that the triplet degeneracy implies  $\operatorname{Re}[F_{\alpha_{2} \alpha_1}(\omega)] = \operatorname{Re}[F_{\alpha_{3} \alpha_1}(\omega)] =\operatorname{Re}[F_{\alpha_{4} \alpha_1}(\omega)]$ explicit evaluation at the eigenenergies of $H_0$ 
gives  $\operatorname{Re}[F_{\alpha_{2,3,4} \alpha_1}(\omega)] = \frac{J(J^2 - \omega^2 + \delta^2)}{[(J + \omega)^2 + \delta^2][(J - \omega)^2 + \delta^2]}$. We therefore use  $ \operatorname{Re}[F_{\alpha_2 \alpha_1}(\omega)]$  as a representative matrix element for  $w^1_{ij}$. The first-order correction matrix thus takes the form

\begin{equation}
w^1_{ij} = \left(\begin{array}{ccc} 0 & -4i \xi e^+\operatorname{Re}[F_{\alpha_2 \alpha_1}(\omega)] &  0 \\ 4i \xi e^+\operatorname{Re}[F_{\alpha_2 \alpha_1}(\omega)]  & 0 & 0\\ 0 & 0 & 0\end{array}\right).
\label{eq:wij1_matrix_Heisenberg_model}
\end{equation}

The total matrix $w_{ij}^0 + w_{ij}^1 $  therefore takes the form

\begin{equation}
w_{ij}^0 + w_{ij}^1 = \left(\begin{array}{ccc} a &   -i \xi  b & 0 \\ i  \xi  b & a & 0\\ 0 &0 & a \end{array}\right)
\label{eq:wij0_wij1}
\end{equation}

where 

\begin{equation}
a=  4 \frac{cosh(\beta J/4)}{Z}
\end{equation}
and

\begin{equation*}
b = 4e^+ \operatorname{Re}[F_{\alpha_2 \alpha_1}(\omega)]
\end{equation*}
\begin{equation}
 =  8 \frac{sinh(\beta J/4)}{Z} \frac{J(J^2 - \omega^2 + \delta^2)}{[(J + \omega)^2 + \delta^2][(J - \omega)^2 + \delta^2]}.
\end{equation}

Diagonalization of Eq. \ref{eq:wij0_wij1} yields eigenvalues $a$, $a - \xi |b|$ and $a + \xi |b|$

\begin{equation}
diag(w_{ij}^0 +  w_{ij}^1) = \left(\begin{array}{ccc} a & 0 & 0 \\ 0 & a- \xi |b| & 0 \\ 0 & 0 & a+  \xi  |b| \end{array}\right).
\end{equation}

The LQU is determined by the maximum eigenvalue [Eq. \ref{eq:LQU_general_2xd}] 

\begin{equation}
u_A (\rho_{AB})= 1 - max \left\lbrace w_{ij} + w_{ij}^1 \right\rbrace,
\end{equation}

which yields the explicit expression

\begin{equation}
    \begin{split}
        u_A(\rho_{AB})& =  1 - 4\left\lbrace  \frac{cosh(\beta J/4)}{Z}  \right.\\
        &\left.  +  2  \xi  \left| \frac{sinh(\beta J/4)}{Z} \frac{J(J^2 - \omega^2 + \delta^2)}{[(J + \omega)^2 + \delta^2][(J - \omega)^2 + \delta^2]}\right| \right\rbrace.
    \end{split}
\label{eq:LQU_total_Heisenberg_model}
\end{equation}

Following an analogous procedure for perturbation along other directions  $\hat{A} = \sigma_x$ and $\sigma_y$, one finds that, although the structure of $w_{ij}^1$ changes, the eigenvalues and the resulting LQU are identical to those obtained for  $\hat{A} = \sigma_z$.  Similarly, applying the local operator to the second spin $S_2$ instead of $S_1$ leaves the LQU unchanged. Both results are a direct consequence of the symmetry of  the Heisenberg model to the spin exchange.

\begin{figure}[hbtp]
\centering
\includegraphics[scale=.11]{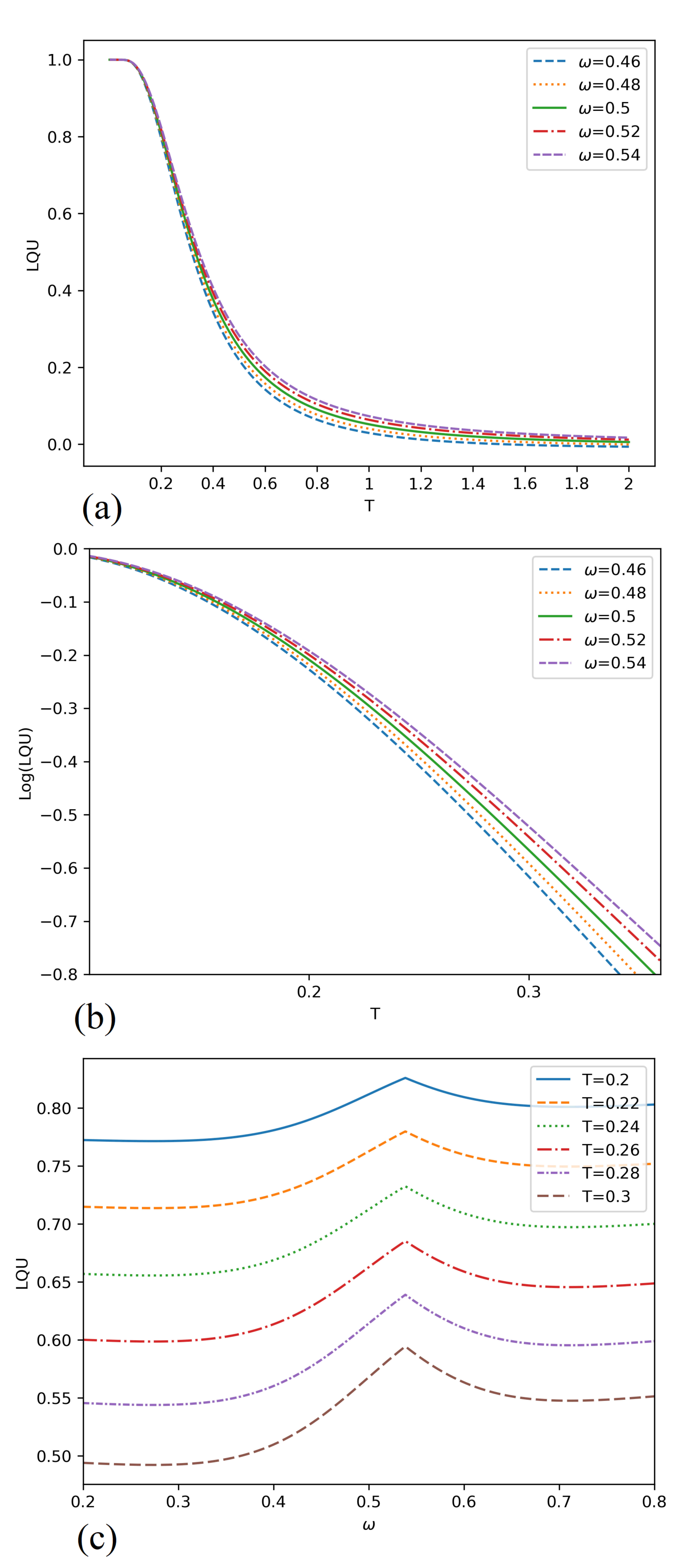}
\caption{
LQU for $\xi = 0.05$, $\delta = 0.2$, and $J = 0.5$.
(a) LQU vs temperature $T$ for driving frequencies $\omega$ near resonance.
(b) $\log(\mathrm{LQU})$, resolving the frequency-induced splitting at higher $T$.
(c) LQU vs $\omega$ for selected $T$, exhibiting resonant enhancement near $\omega = J$.
}
\label{fig:LQU_and_Log(LQU)_of_T}
\end{figure}

Figure ~\ref{fig:LQU_and_Log(LQU)_of_T}(a) shows the LQU as a function of temperature for a range of driving frequencies in the vicinity of the resonance $\omega = J = 0.5$, as given by Eq.~(\ref{eq:LQU_total_Heisenberg_model}). The LQU decreases monotonically with increasing temperature. At low temperatures, the frequency dependence is weak, whereas it becomes increasingly pronounced at higher temperatures. To better resolve the differences among frequency values and the dynamical induced discord in the linear response regime, the logarithm of the LQU is plotted in Fig. ~\ref{fig:LQU_and_Log(LQU)_of_T}(b). At low temperatures, the LQU approaches its maximum, consistent with the unperturbed result of Eq.~(\ref{eq:LQU_w0_Heisenberg_model}). As the temperature increases, the curves separate, revealing a clear frequency-dependent modulation of the nonclassicality induced by the external perturbation; curves corresponding to frequencies near resonance exhibit the largest LQU values.  The effect of frequency is shown in Figure~\ref{fig:LQU_and_Log(LQU)_of_T}(c) , where  we present the LQU as a function of frequency for several temperatures. A resonant enhancement of nonclassicality is clearly  observed near $\omega = J = 0.5$
 for all temperatures, with the LQU approaching its maximum near unity at the lowest temperature shown. The role of the broadening and perturbation parameters is also evident: increasing $\delta$
 shifts the LQU peak toward higher frequencies (for $\delta = 0$ the maximum occurs precisely at $\omega = J$), while increasing $\xi$ enhances the amplitude of the peak.

\begin{figure}[hbtp]
\centering
\includegraphics[scale= .11]{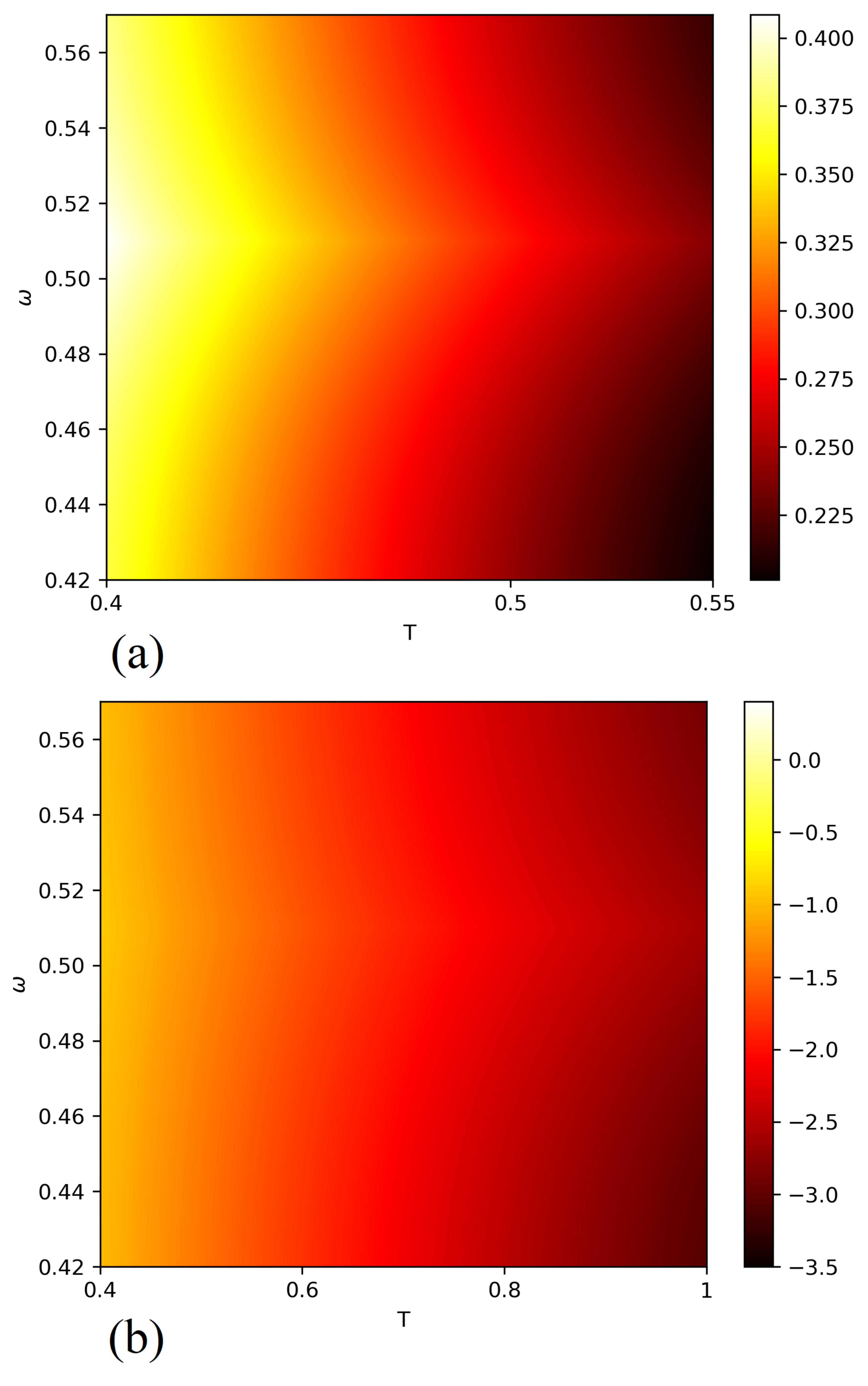}
\caption{
LQU in the $(T,\omega)$ plane for $\xi = 0.05$, $\delta = 0.2$, and $J = 0.5$.
(a) LQU.
(b) $\log(\mathrm{LQU})$, highlighting the resonant ridge near $\omega = J$ and its persistence with increasing temperature.
}
\label{fig:LQU_and_log(LQU)_color_map}
\end{figure}

Figures ~\ref{fig:LQU_and_log(LQU)_color_map}(a) and \ref{fig:LQU_and_log(LQU)_color_map}(b) display color maps of the LQU and its logarithm as simultaneous functions of frequency $\omega$
and temperature $T$. In both panels, a clear enhancement of nonclassicality is evident near $\omega = J = 0.5$, which persists --- though it gradually attenuates --- as the temperature increases.

To determine whether the dynamical induced observed nonclassicality is of  pure quantum-discord type or involves entanglement, we compute the concurrence ~\cite{ hill1997entanglement}  (entanglement measurement) of the perturbed state in the linear regime  $\rho = \rho_0 + \epsilon \rho_1$ and compare it with the  obtained LQU.

To compute the concurrence of the perturbed state $\rho = \rho_0 + \epsilon\rho_1$, it is necessary to evaluate the full density matrix in the computational basis. In the linear response regime, the matrix elements of $\rho$  acquire an imaginary contribution through $\operatorname{Im}[F_{\alpha_m \alpha_l}]$, given by

\begin{equation}
    \begin{split}
        &\operatorname{Im}[F_{\alpha_m \alpha_l}(\omega)]  \\
        & =\left\lbrace\frac{ 2 \delta [(E_{\alpha_m} - E_{\alpha_l})^2 + \omega^2 + \delta^2]}{[(E_{\alpha_m}-E_{\alpha_l}+\omega)^2 + \delta^2][(E_{\alpha_m}-E_{\alpha_l}-\omega)^2 + \delta^2] }\right\rbrace.
    \end{split}
\end{equation}

With the same index ordering $l=1$ and $m \in \{2,3,4\}$ as in Eq.(\ref{eq:ReF}), the triplet degeneracy implies  $\operatorname{Im}[F_{\alpha_2 \alpha_1}] =\operatorname{Im}[F_{\alpha_3 \alpha_1}]=\operatorname{Im}[F_{\alpha_4 \alpha_1}]$ and explicit evaluation at the eigenenergies of $H_0$ gives $\operatorname{Im}[F_{\alpha_{2,3,4} \alpha_1}]= 2 \delta\left\lbrace\frac{J^2 + \omega^2 + \delta^2}{[(J+\omega)^2 + \delta^2][(J-\omega)^2 + \delta^2]}\right\rbrace$. We therefore use $\operatorname{Im}[F_{\alpha_2 \alpha_1}])$ as the representative matrix element entering the concurrence expression, and define for convenience  $\Omega \equiv \sqrt{1 + 4 \xi^2 \operatorname{Im}[F_{\alpha_n \alpha_m}]^2} $, which will appear explicitly in both the concurrence and the critical temperature. With this result, the density matrix in the computational basis takes the form 

\begin{equation}
\rho = D \left(\begin{array}{cccc} A & 0 & 0 & 0 \\ 0 & B_+ & C & 0\\  0 &  C^* & B_- & 0  \\ 0 & 0 & 0 & A \end{array}\right)
\label{eq:computational_basis_density_matrix}
\end{equation}

where $A= e^{\beta J/2}$, $B_{+,-}=cosh(\beta J/2) \pm  \xi sinh(\beta J/2) \operatorname{Re}[F_{\alpha_2 \alpha_1}]$, $C= sinh(\beta J/2)(1 + 2i\operatorname{Im}[F_{\alpha_2 \alpha_1}])$ and $D= \frac{e^{\beta J/4}}{Z}$. From the structure of Eq. \ref{eq:computational_basis_density_matrix}, of X form the concurrence is

\begin{equation}
\mathcal{C}= 2max(0,|C|-A),
\end{equation}

which yields the explicit expression

\begin{equation}
        \mathcal{C} = \left|\frac{e^{3/4 \beta J} - e^{-1/4 \beta J}}{Z} \right| \Omega  - 2 \frac{e^{-1/4 \beta J}}{Z}.
\label{eq:concurrence}
\end{equation}

\begin{figure}[hbtp]
\centering
\includegraphics[scale=.11]{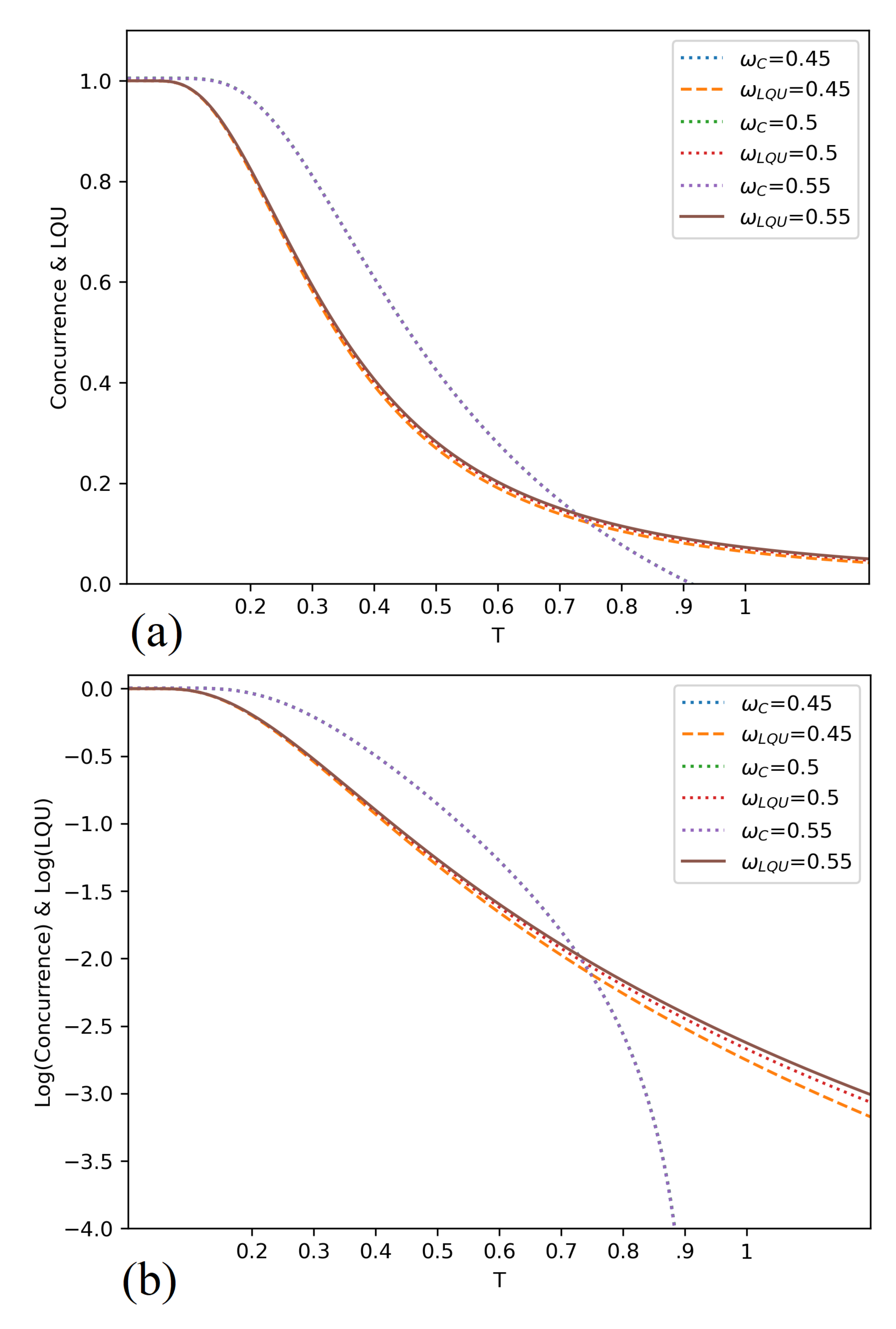}
\caption{
LQU and concurrence for $\xi = 0.01$, $\delta = 0.2$, and $J = 0.5$.
(a) Temperature dependence for several $\omega$.
(b) Logarithmic scale showing finite LQU above the temperature where concurrence vanishes, evidencing field-induced discord without entanglement.
}
\label{fig:concurrence_and_LQU}
\end{figure}

With both the LQU Eq.~(\ref{eq:LQU_total_Heisenberg_model}) and the concurrence, Eq.~(\ref{eq:concurrence}), in hand, we now compare the two measures directly to determine the nature of the field-induced nonclassicality. Figure~\ref{fig:concurrence_and_LQU} shows (a) the LQU and concurrence and (b) their logarithms, as functions of temperature for several driving frequencies. These plots focus on the temperature range above which the concurrence vanishes. A frequency-dependent splitting of the LQU curves is observed, indicating an enhancement of nonclassicality induced by the external field. This directly demonstrates the perturbative effect discussed in connection with Figs.~\ref{fig:LQU_and_Log(LQU)_of_T}~(a) and~(b):  above the entanglement threshold, where $ \mathcal{C} = 0$, the enhancement of nonclassical correlations is of purely quantum-discord type rather than entanglement, and is frequency dependent. The LQU remains finite and exhibits a clear frequency modulation well above the temperature at which the concurrence vanishes — a signature of field-induced quantum discord within the linear response framework.

\begin{figure}[hbtp]
\centering
\includegraphics[scale=.11]{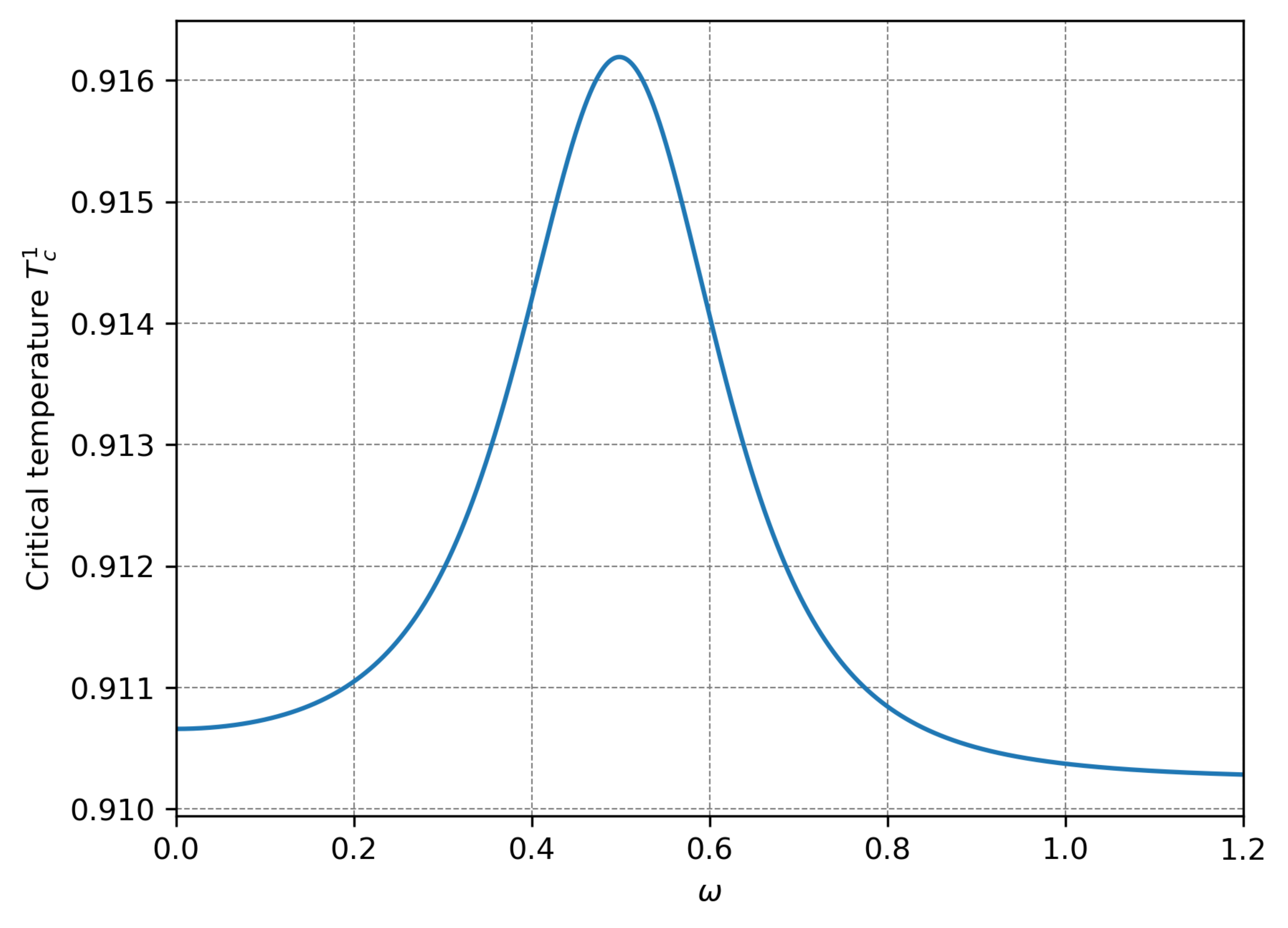}
\caption{
Critical temperature $T_c^{(1)}$ vs driving frequency $\omega$ for $\xi = 0.01$, $\delta = 0.2$, and $J = 0.5$.
The weak frequency dependence indicates that entanglement is largely insensitive to the drive, in contrast to the resonant behavior of the LQU.
}
\label{fig:critical_temperature}
\end{figure}

The critical temperature $T_c$  for the onset of a separable state is determined by the condition  $\mathcal{C}= 0$ \cite{fu2004critical}. In the absence of the perturbation, this gives $T_c^0 = \frac{J}{\ln(3)}$, while in the presence of the perturbation of strength $\xi$  one obtains  $T_c^1 = \frac{J}{\ln\left(\frac{\Omega + 2}{\Omega}\right)}$. Figure \ref{fig:critical_temperature}  displays $T_c^1$ as a function of frequency. Only a marginal increase is observed near resonance, indicating that entanglement is essentially insensitive to the driving frequency. We therefore conclude that the frequency-dependent enhancements of nonclassicality observed above $T_c^0$ are of purely quantum-discord type, with frequency acting as a tunable modulator of their amplitude.

These results demonstrate the utility of the LQU formalism in the linear response regime and establish a foundation for its extension to a broader class of quantum systems in future work. In particular, one can consider systems whose unperturbed state has vanishing quantum discord, and for which the external perturbation within the linear response regime induces a transition to states with finite nonclassicality.

\section{Conclusion}
\label{sec:summary}
We have presented a systematic general  perturbation theory for the local quantum uncertainty
(LQU) applicable to composite quantum systems of arbitrary dimension $d_1 \times d_2$.
The central result is a first-order expansion of the density matrix square root
$\rho^{1/2}$, derived via an integral representation based on the gamma function,
which allows the LQU optimization to be reduced to the diagonalization of a
$(d_1^2-1)\times(d_1^2-1)$ matrix $w = w^0+ w^1$
expressed in the $\mathrm{SU}(d_1)$ generator basis. The equilibrium contribution
$w^0$ recovers the known closed-form result for unperturbed 
states~~\cite{closedformwang}, while the perturbative correction $w^1$ provides
a general and direct method to evaluate the modification of nonclassicality induced
by an arbitrary perturbation $\rho_1$, given solely in terms of the eigenstates and
eigenvalues of the unperturbed density matrix $\rho_0$.

The theory was further developed within the quantum linear response regime, where
the perturbation arises from a time-dependent external peridic field $f(t)$ coupled to the
system via an operator $\hat{A}$. In this context, the perturbed density matrix
$\rho_1$ acquires an explicit dependence on the driving frequency $\omega$, the
energy spectrum of the equilibrium Hamiltonian $H_0$, the thermal occupation
probabilities $\lambda_{\alpha_n}$, and the matrix elements $\braket{ \alpha_n |
\hat{A} | \alpha_m}$. This structure enables a systematic investigation of
field-induced quantum discord across a broad class of physical systems, including
quantum spin arrays with isotropic, anisotropic, Dzyaloshinskii--Moriya, and
Kaplan--Shekhtman--Entin-Wohlman--Aharony interactions, as well as atoms coupled
to cavity fields and optomechanical systems.

As an analytically tractable application, the formalism was implemented for the
isotropic Heisenberg model of two coupled spins driven by a local periodic magnetic
field. Closed-form expressions for the LQU were obtained as explicit functions of
temperature $T$ and driving frequency $\omega$. The matrix $w^0$ was
found to be proportional to the identity, reflecting the isotropy of the model,
while $w^1$ develops nonzero off-diagonal elements in the $xy$ sector,
breaking the degeneracy and introducing a frequency-dependent splitting of the
eigenvalues. The resulting LQU exhibits a resonant enhancement near $\omega = J$,
whose amplitude is controlled by the perturbation strength $\xi$ and whose width
is governed by the broadening parameter $\delta$.

Comparison with the concurrence of the perturbed state reveals a qualitative
distinction between the two regimes of nonclassicality. Below the critical
temperature $T_c^0 = J/\ln(3)$, both entanglement and quantum discord coexist,
with the latter persisting and exhibiting frequency modulation. Above $T_c$,
the concurrence vanishes identically while the LQU remains finite and continues
to display a clear resonant structure as a function of $\omega$. The critical
temperature in the presence of the perturbation, $T_c^1 = J/\ln[(\Omega+2)/\Omega]$,
shows only a minimal shift near resonance, confirming that the frequency-dependent
enhancement of nonclassicality observed at high temperatures is of purely
quantum-discord type and is not accompanied by the generation of entanglement.
This result demonstrates that an external periodic field can act as a tunable
modulator of discord-type correlations without altering the entanglement structure
of the system.

The framework developed here is not restricted to spin models. Its generality
makes it directly applicable to any bipartite system admitting a well-defined
equilibrium Hamiltonian $H_0$ and a perturbation expressible in the linear
response form $H_1 = -\hat{A}f(t)$. In particular, extensions to higher-dimensional
subsystems with $d_1 > 2$, multipartite configurations, and non-equilibrium
steady states described by Lindblad master equations are natural directions for
future work. The connection between the perturbative LQU and experimentally
accessible response functions, such as the dynamic susceptibility, also suggests
possible routes toward the direct measurement of discord-type correlations in
driven quantum systems.

\section*{Acknowledgments}

A.A.J. acknowledges support from SECIHTI. The authors also acknowledge partial financial support from DGAPA-UNAM-PAPIIT, Grant No. IN111122 (Mexico).

\section{Appendix}

\subsection{Perturbative expansion of the density matrix square root}
\label{apen:rho^1/2}

To avoid notational conflict with the perturbed density matrix $\rho = \rho_0 + \epsilon \rho_1$ introduced in the main text, we temporarily denote the argument of the square-root map as $\gamma$. We compute the first-order operator derivative of  $\gamma^{1/2}$ to construct a Taylor-like expansion, using the Stieltjes integral representation

\begin{equation}
\gamma^{1/2} = \frac{sen(\pi/2)}{\pi} \int_0^{\infty} \frac{\gamma}{\gamma+ t}t^{-1/2}dt.
\end{equation}

Denoting the perturbation as

$$\epsilon \rho_1 \equiv \nu $$

we introduce the operators

\begin{equation}
\gamma_+ = (\rho_0 + \epsilon  \nu) 
\end{equation}

and

\begin{equation}
\gamma_0 = \rho_0,
\end{equation}

 from which the first-order derivative is determined as
 
\begin{equation}
\left. \frac{d}{d\epsilon}\right|_{\epsilon=0} \gamma^{1/2}  = \lim_{\epsilon \rightarrow 0} \frac{\gamma_+^{1/2} - \gamma_0^{1/2}}{\epsilon}.
\end{equation}

Expressing this derivative in integral form via the Stieltjes representation and expanding the integrand, one obtains

\begin{equation}
    \begin{split}
        \int \frac{\gamma_+}{\gamma_+ + t} - \frac{\gamma_0}{\gamma_0 + t} t^{-1/2} dt = \\
        \int \frac{( \gamma_+ \gamma_0 + \gamma_+ t - \gamma_0 \gamma_+ - \gamma_0 t)t^{-1/2}}{(\gamma_+ +t)(\gamma_0 + t)} t^{-1/2}dt.
    \end{split}
\label{int_deriv1}
\end{equation}

Substituting the explicit forms of $\gamma_+$ and $\gamma_0$ into  Eq.~(\ref{int_deriv1}) and taking the limit $\epsilon \rightarrow 0 $, we obtain

\begin{equation}
\frac{\partial(\gamma^{1/2})}{\partial\epsilon} = \lim_{\epsilon \rightarrow 0} \frac{1}{\epsilon \pi} \int \frac{\epsilon [\nu, \rho_0] + \epsilon \nu t}{(\rho_0 + \epsilon \nu + t)(\rho_0 + t)}t^{-1/2}dt.
\end{equation}

Assembling the zeroth- and first-order contributions, the full operator integral expression for $\rho^{1/2}$ reads

\begin{equation}
    \begin{split}
        \rho^{1/2} =  \frac{1}{\pi} \int \frac{\gamma}{\gamma + t}t^{-1/2}dt\\
        +  \lim_{\epsilon \rightarrow 0} \frac{1}{\epsilon \pi} \int \frac{\epsilon [\nu, \rho_0] + \epsilon \nu t}{(\rho_0 + \epsilon \nu + t)(\rho_0 + t)}t^{-1/2}dt. 
    \end{split}
\end{equation}

The first term of the integral expression, written in the $H_0$ eigenbasis using the spectral decomposition $\rho_0 = \sum_i \lambda_i \ket{\psi_i}\bra{\psi_i}$, is given by

\begin{equation}
    \begin{split}
        \sum_{i j} \lambda_i\ket{\psi_i}\bra{\psi_i}(\rho_0 + t)^{-1}\ket{\psi_j}\bra{\psi_j}t^{-1/2}dt\\
        = \sum_{\psi_i} \int \frac{\lambda_i t^{-1/2}}{\lambda_i + t} dt \ket{\psi_i}\bra{\psi_i}.
    \end{split}
\end{equation}

The second term, which encodes the first-order response of the square root to the perturbation $\nu$, evaluates to

\begin{equation}
    \begin{split}
        \sum_{i,j,k,l} \int \ket{\psi_i}\bra{\psi_i}(\rho_0 + t)^{-1}\ket{\psi_j}\\
        \times \bra{\psi_j} \nu \ket{\psi_k}\bra{\psi_k}(\rho_0 + t)^{-1}\ket{\psi_l}\bra{\psi_l} t^{1/2}dt\\
        = \sum_{i j} \int \frac{\nu_{ij}t^{1/2}}{(\lambda_i + t)(\lambda_j+t)}\ket{\psi_i}\bra{\psi_j}dt.
    \end{split}
\end{equation}

Restoring the original notation  $\nu = \epsilon \rho_1$ and collecting both contributions under a common integral, we arrive a

\begin{equation}
    \begin{split}
        \pi \rho^{1/2} = \sum_i \lambda_i \int_0^{\infty} \frac{t^{-1/2}}{\lambda_i + t} dt \ket{\psi_i}\bra{\psi_i} \\
        + \epsilon \sum_{ij} \rho_{1 ij} \int_0^{\infty} \frac{t^{1/2}}{(\lambda_i + t)(\lambda_j + t)} dt  \ket{\psi_i}\bra{\psi_j}.
    \end{split}
\end{equation}

Evaluating the resulting integrals using standard Stieltjes identities yields Eq.~(\ref{eq:rho1/2_general}) of the main text

\begin{equation}
    \begin{split}
        \rho^{1/2} = \sum_{i}\lambda_{i}^{1/2}\ket{\psi_i}\bra{\psi_i} \\
        + \epsilon \sum_{ij}\rho_{1ij} \left( \frac{\lambda^{1/2}_{i} -\lambda^{1/2}_{j}}{\lambda_{i} -\lambda_{j}} \right)\ket{\psi_i}\bra{\psi_j}\\
        + O(\epsilon^2).
    \end{split}
    \label{eq:app_rho^1/2}
\end{equation}

The result in Eq.~(\ref{eq:app_rho^1/2}) provides the explicit form of $\rho^{1/2}$ required to evaluate the skew information in Sec.\ref{sec:LQU-PT}, and constitutes the key analytical ingredient of the perturbation theory developed in this work.

\subsection{Matrix elements of \texorpdfstring{$w_{ij}^0$}{wij0} and \texorpdfstring{$w_{ij}^1$}{wij1} for the isotropic Heisenberg model}
\label{apen:wij}

We evaluate the matrix elements of the local Pauli operators in the eigenbasis $\ket{\alpha_n}$ of $H_0 = J S_1 \cdot S_2$. For each component $i = x, y, z$, the nonvanishing matrix elements $\braket{\alpha_n|\sigma_i \otimes \mathbb{I}_B| \alpha_m}$ are

\begin{equation}
    \begin{split}
        \braket{\alpha_1|\sigma_x \otimes I_B | \alpha_{3}}=  \frac{1}{\sqrt{2}}, \braket{\alpha_3|\sigma_x \otimes I_B | \alpha_{1}}=  \frac{1}{\sqrt{2}}\\
        \braket{\alpha_1|\sigma_x \otimes I_B | \alpha_{4}}=  -\frac{1}{\sqrt{2}}, \braket{\alpha_4|\sigma_x \otimes I_B | \alpha_{1}}=  -\frac{1}{\sqrt{2}}\\
        \braket{\alpha_2|\sigma_x \otimes I_B | \alpha_{3}}=  \frac{1}{\sqrt{2}}, \braket{\alpha_3|\sigma_x \otimes I_B | \alpha_{2}}=  \frac{1}{\sqrt{2}}\\
        \braket{\alpha_2|\sigma_x \otimes I_B | \alpha_{4}}=  \frac{1}{\sqrt{2}}, \braket{\alpha_4|\sigma_x \otimes I_B | \alpha_{2}}=  \frac{1}{\sqrt{2}}
    \end{split}
    \label{eq:app_B_matrix_elements_x}
\end{equation}

when $i= \sigma_x$,

\begin{equation}
\begin{split}
    \braket{\alpha_1|\sigma_y \otimes I_B | \alpha_{3}}=  -\frac{i}{\sqrt{2}}, \braket{\alpha_3|\sigma_y \otimes I_B | \alpha_{1}}= \frac{i}{\sqrt{2}}\\
    \braket{\alpha_1|\sigma_y \otimes I_B | \alpha_{4}}=  -\frac{i}{\sqrt{2}}, \braket{\alpha_4|\sigma_y \otimes I_B | \alpha_{1}}=  \frac{i}{\sqrt{2}}\\
    \braket{\alpha_2|\sigma_y \otimes I_B | \alpha_{3}}=  \frac{-i}{\sqrt{2}}, \braket{\alpha_3|\sigma_y \otimes I_B | \alpha_{2}}=  \frac{i}{\sqrt{2}}\\
    \braket{\alpha_2|\sigma_y \otimes I_B | \alpha_{4}}=  \frac{i}{\sqrt{2}}, \braket{\alpha_4|\sigma_y \otimes I_B | \alpha_{2}}=  -\frac{i}{\sqrt{2}},
\end{split}
\label{eq:app_B_matrix_elements_y}
\end{equation}

when $i= \sigma_y$, and

\begin{equation}
    \begin{split}
        \braket{\alpha_1|\sigma_z \otimes I_B | \alpha_{2}}=  1, \braket{\alpha_2|\sigma_z \otimes I_B | \alpha_{1}}= 1 \\
        \braket{\alpha_3|\sigma_z \otimes I_B | \alpha_{3}}=  -1, \braket{\alpha_4|\sigma_z \otimes I_B | \alpha_{4}}=1
    \end{split}
    \label{eq:app_B_matrix_elements_z}
\end{equation}

when $i= \sigma_z$.

Inspection of Eqs.~(\ref{eq:app_B_matrix_elements_x})–(\ref{eq:app_B_matrix_elements_z}) reveals that the products $\braket{\alpha_n|\sigma_i \otimes \mathbb{I}_B| \alpha_m}\braket{\alpha_m|\sigma_j \otimes \mathbb{I}_B| \alpha_n}$ vanish upon summation for $i = x$, $j = y$ and $i = y$, $j = x$, owing to the alternating sign pattern of the matrix elements. Analogously, all cross-combinations with $i \in \{x,y\}$ and $j=z$, or vice versa, vanish by symmetry. Consequently, all off-diagonal elements of $w_{ij}^0$ are zero. Carrying out the explicit sum for $i = j = x$, using the matrix elements listed in Eq.~(\ref{eq:app_B_matrix_elements_x}), one finds

\begin{equation}
    \begin{split}
        w_{xx}^0 =  \sum_{\alpha_n \alpha_m} \lambda_{\alpha_n}^{1/2}\lambda_{\alpha_m}^{1/2}\braket{ \alpha_n|\sigma_x \otimes I_B|\alpha_m}\braket{ \alpha_m| \sigma_x \otimes I_B|\alpha_n}\\
        = \frac{1}{2} \frac{e^{3/8 \beta J}}{Z^{1/2}}\frac{e^{-1/8 \beta J}}{Z^{1/2}} + \frac{1}{2} \frac{e^{3/8 \beta J}}{Z^{1/2}}\frac{e^{-1/8 \beta J}}{Z^{1/2}}\\ + \frac{1}{2} \frac{e^{-1/8 \beta J}}{Z^{1/2}}\frac{e^{-1/8 \beta J}}{Z^{1/2}}  + \frac{1}{2} \frac{e^{-1/8 \beta J}}{Z^{1/2}}\frac{e^{-1/8 \beta J}}{Z^{1/2}}\\ + \frac{1}{2} \frac{e^{3/8 \beta J}}{Z^{1/2}}\frac{e^{-1/8 \beta J}}{Z^{1/2}} + \frac{1}{2} \frac{e^{-1/8 \beta J}}{Z^{1/2}}\frac{e^{-1/8 \beta J}}{Z^{1/2}}\\ + \frac{1}{2} \frac{e^{3/8 \beta J}}{Z^{1/2}}\frac{e^{-1/8 \beta J}}{Z^{1/2}} + \frac{1}{2} \frac{e^{3-1/8 \beta J}}{Z^{1/2}}\frac{e^{-1/8 \beta J}}{Z^{1/2}}\\
        =2\frac{e^{1/4 \beta J}}{Z} + 2\frac{e^{-1/4 \beta J}}{Z}\\
        = 4\frac{cosh(\beta J/4)}{Z}).\\
    \end{split}
\end{equation}

An analogous calculation for $i = j = y, z$ yields $ w_{yy}^0 = w_{zz}^0$, reflecting the spin-rotation symmetry of the isotropic Heisenberg model. The matrix $w_{ij}^0$ is therefore proportional to the identity

\begin{equation}
w_{ij}^0 = 4\frac{cosh(\beta J/4)}{Z} \left(\begin{array}{ccc} 1 & 0 & 0 \\ 0 & 1 & 0\\ 0 & 0 & 1\end{array}\right).
\end{equation}

Turning to the first-order correction $w_{ij}^1$, we note that the factor $(\lambda_{\alpha_m}^{1/2} - \lambda_{\alpha_l}^{1/2})$  in Eq.~(\ref{eq:wij_Heisenberg_model}) suppresses all diagonal contributions in which $m = l$. Combined with the antisymmetry relation  $ F_{\alpha_m \alpha_l}(\omega) = - F^*_{\alpha_l \alpha_m}(\omega)$, this restricts the nonvanishing elements of  $w_{ij}^1$ to the off-diagonal sector $i = x$, $j = y$ (and its transpose). Evaluating the surviving terms in Eq.~(\ref{eq:wij_Heisenberg_model}) for the index pair $(i,j) = (y,x)$ gives

\begin{equation}
    \begin{split}
        w^1_{yx}  =   -2 \xi  \sum_{\alpha_n \alpha_m \alpha_l} \lambda_{\alpha_n}^{1/2}(\lambda_{\alpha_m}^{1/2} - \lambda_{\alpha_l}^{1/2}) \braket{\alpha_m|\sigma_z \otimes \mathbb{I}_{B}|\alpha_l}\\
        \times F_{\alpha_m \alpha_l}(\omega) \braket{ \alpha_n|\sigma_x \otimes \mathbb{I}_{B}|\alpha_m} \braket{ \alpha_l| \sigma_y \otimes \mathbb{I}_{B}|\alpha_n}\\
        = -2 \xi (\frac{i}{2}e^+ F_{12}  -\frac{i}{2}e^+ F_{21} -\frac{i}{2}e^+ F_{21} + \frac{i}{2}e^+ F_{12})\\
        = -2 \xi( -\frac{i}{2}e^+ F^*_{21}  -\frac{i}{2}e^+ F_{21} -\frac{i}{2}e^+ F_{21} - \frac{i}{2}e^+ F^*_{21})\\
        = 4 i\xi e^+ Re[F_{21}].
    \end{split}
\end{equation}
\vspace{-0.5em}

The Hermiticity of $w^1_{ij}$ requires $ w^1_{xy}   = (w^{1}_{yx})^*$ , which together with the explicit evaluation above yields the matrix

\begin{equation}
w^1_{ij} = \left(\begin{array}{ccc} 0 & -4i \xi e^+Re[F_{\alpha_2 \alpha_1}(\omega)] &  0 \\ 4i \xi e^+Re[F_{\alpha_2 \alpha_1}(\omega)]  & 0 & 0\\ 0 & 0 & 0\end{array}\right).
\end{equation}

The matrix elements computed above confirm that the spin symmetry of the Heisenberg model is preserved at the level of $w^0_{ij}$, and is broken only by the off-diagonal structure of $w^1_{ij}$ introduced by the external perturbation.

\newpage

\bibliography{bibl_art} 

\end{document}